\documentclass[10pt]{iopart}
\pdfoutput=1
\usepackage{graphicx}
\usepackage{setstack}
\usepackage{iopams}
\usepackage[T1]{fontenc}
\usepackage[latin9]{inputenc}
\usepackage{bm}
\usepackage{color}
\usepackage{comment}
\usepackage{hyperref}
\begin{document}
\title{Absorption of single-layer hexagonal boron nitride in the ultraviolet}
\author{J. C. G. Henriques$^{1}$, G. B. Ventura$^{2}$, C. D. M. Fernandes$^{1}$,
and N. M. R. Peres$^{1,3}$}
\ead{peres@fisica.uminho.pt}

\address{$^{1}$Department and Centre of Physics, and QuantaLab, University
of Minho, Campus of Gualtar, 4710-057, Braga, Portugal}

\address{$^{2}$Department of Physics and Astronomy, Faculty of Sciences of
the University of Porto, Rua Campo Alegre 687, 4169-007, Porto, Portugal}

\address{$^{3}$International Iberian Nanotechnology Laboratory (INL), Av.
Mestre José Veiga, 4715-330, Braga, Portugal}

\date{\today}

\begin{abstract}
In this paper we theoretically describe the absorption of hexagonal
boron nitride (hBN) single layer. We develop the necessary formalism
and present an efficient method for solving the Wannier equation for
excitons. 
We give predictions for the absorption of hBN on quartz and
on graphite. We compare our predictions with recently published results
[Elias {\it et al.}, Nat. Comm. {\bf 10}, 2639  (2019)] for a
monolayer of hBN on graphite.
The spontaneous radiative lifetime of excitons in hBN is
also computed. 
We argue that the optical properties of hBN in the
ultraviolet are very useful for the study of peptides and other biomolecules.
\end{abstract}

\submitto{\TDM}
\maketitle
\ioptwocol

\section{Introduction}
\label{sec:Intro}

Cyclic $\beta-$helical peptides are biomolecules with an intense absorption around $\sim200-240$ nm, that is, in the ultraviolet \cite{Dmitri2013} (see also \cite{Eigil1967,Bode1996}). They play an important role in the biophysics of the human body and their optical activity has been studied when deposited in quartz, as this latter material is transparent in the ultraviolet. Incidentally, peptides
are playing a role  in some unsuspected applications \cite{Berger2015}.
Due to the biological relevance of peptides, it would be both interesting and useful to have a material with a strong optical response in the same spectral range. It would be specially interesting if this sought material could have a flat surface and could support polaritons with an electric field decaying exponentially away from its surface. The combination of the two characteristics allows, in princple, the construction of a sensor that is capable of detecting minute quantities of peptides. It has been shown that hexagonal boron nitride (hBN) deposited on quartz has a strong resonance around this frequency of interest \cite{Watanabe2004,Arnaud2006,Marini2008,Song2010,Stehle2015,Cao2013,Cassabois2016}; the nature of which is explained by the formation of a bound electron-hole pair 
--an exciton-- when electromagnetic radiation of a certain frequency shines on the material. Quartz is itself a widely used dielectric in electronics and is employed in graphene-based transistors for the detection of biomolecules \cite{Dontschuk2015,Pedro2019}. The system that we are considering in this paper is thus easily available for further studies by the community. Though it would seem like a coincidence that the excitonic response of hBN on quartz matches the optical response of peptides on the same material, it must be pointed out that the excitonic resonance depends heavily on the strength of the electron-electron interactions, which are screened differently by different substrates. This opens the possibility for the tuning of the spectral position of the excitonic resonance by means of a judicious choice of the substrate. Additionally, an external magnetic field can be added to the setup, which would allow for an additional degree of freedom available in peptides (their chirality) to be explored.

Although the bibliography on excitons in transition metal dichalcogenides (TMD) is vast and diverse \cite{Heinz2018}, the same does not apply to hBN \cite{Zhang2017}. This material has been mainly
envisioned as an ultra-flat substrate for graphene- or TMD-based electronic and optoelectronic devices \cite{Wang2019} and, although this utility is certainly of importance, we will show here that the optical response of hBN in the ultraviolet is interesting \emph{per se}. Hexagonal boron nitride is also known for its hyperbolic nature in the infrared \cite{Low2015,Kostya2015,Low2017}, allowing for the existence of focused guided modes in hBN slabs \cite{Xu2017,Dai2014}. This characteristic, also interesting in itself, is however immaterial for the spectral range we are exploring in this paper. The excitonic response of suspended single-layer hBN has recently been studied using \emph{ab-initio} and model-Hamiltonian approaches \cite{Ferreira2019}. The latter approach is particularly useful since it strips the problem from all its supplemental details and allows us to focus on the more fundamental aspects. Often, the approach based on model Hamiltonians allows for analytical solutions and the immediate insights that this brings.

\begin{figure}
\includegraphics[scale=0.4]{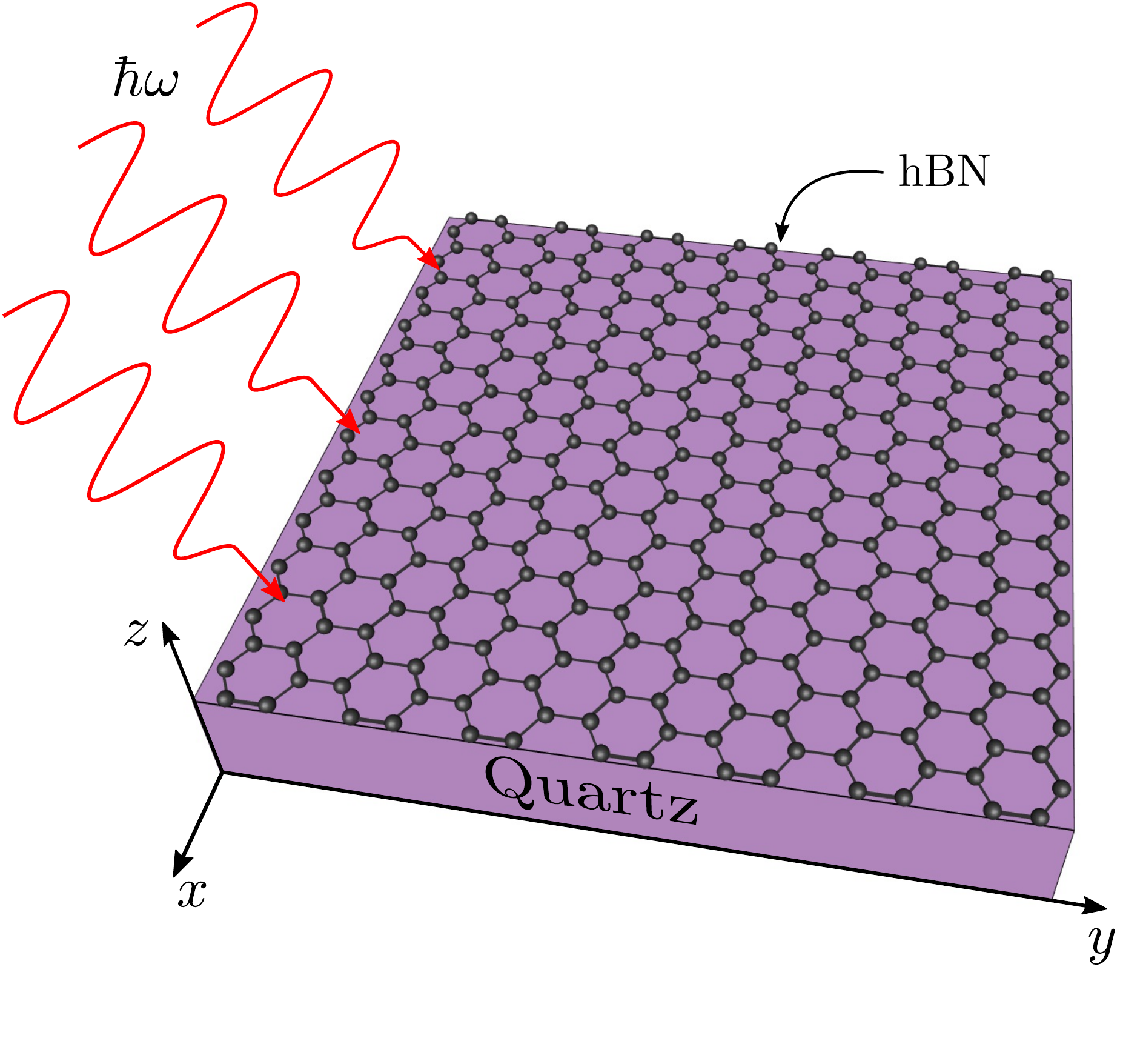}

\includegraphics[scale=0.5]{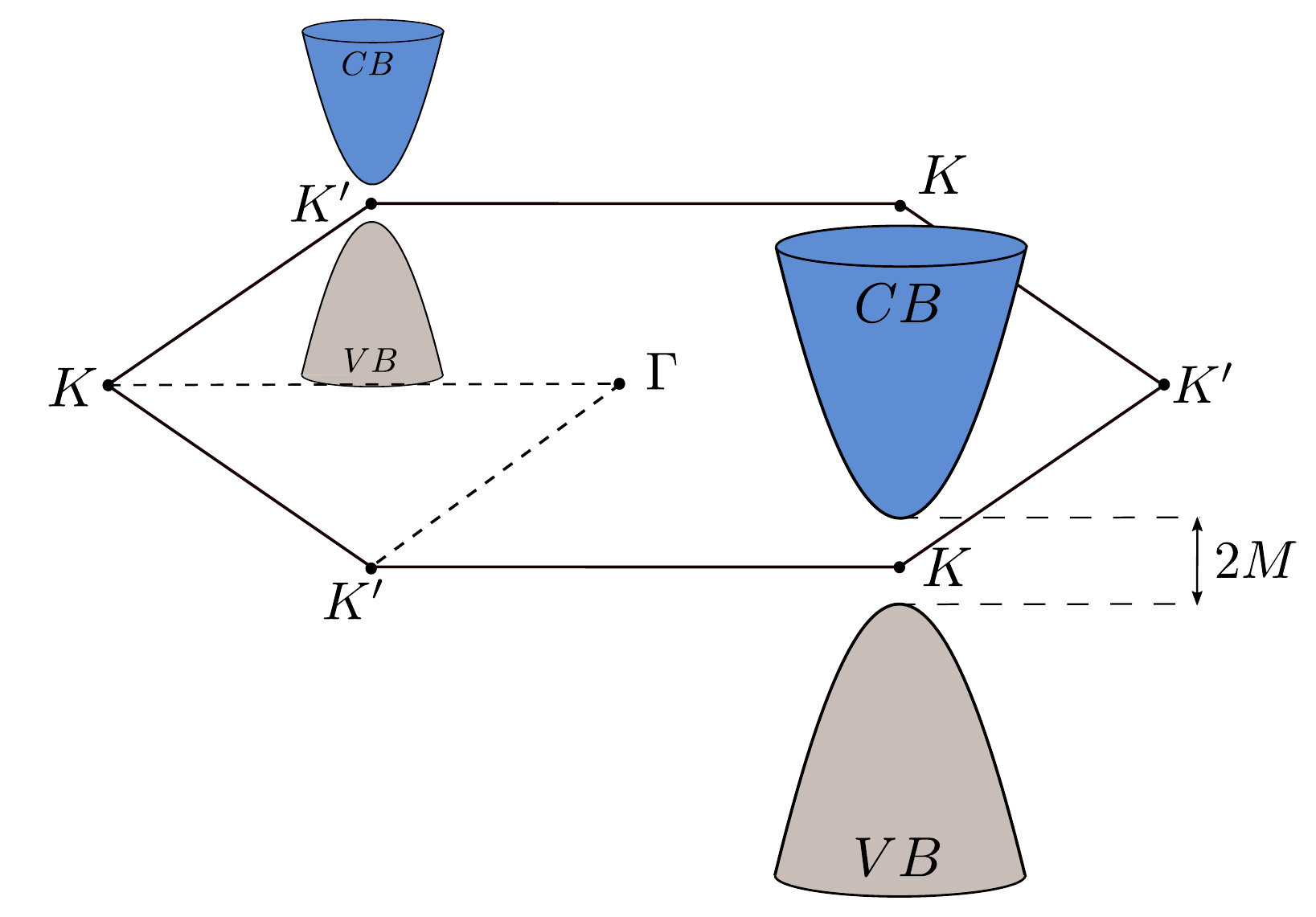}

\caption{Representation of the experimental setup and Brillouin zone of hBN.
The top image shows the experimental setup:  a hBN layer
on quartz with photons shining on the system. On the bottom image
we represent the hexagonal Brillouin zone of hBN together with the
gapped Dirac cones existing at the edges of the zone; the \emph{bare}
(without exchange corrections) energy gap is $2M$.\label{fig:Representation-of-the}}

\end{figure}

Hexagonal boron nitride has a hexagonal lattice, much like graphene, with two atoms per unit cell, a nitrogen and a boron. The two nonequivalent sites in the unit cell means that the electrons hopping around the 2D lattice can actually distinguish whether they are sitting on a boron or on a nitrogen atom. This property, contrary to graphene, leads to the opening of an energy gap at the corners of the Brillouin zone, so its dispersion relation is no-longer  conic. As a consequence, the electronic motion is described by a massive Dirac equation \cite{Hunt2013,Andre2017} in two-dimensions (2D), whose energy gap reads
$2M$. From the relativistic dispersion relation 
 an effective mass reading $m^{\ast}=M/v_{F}^{2}$ can be retrieved, where $2M$  is determined by fitting the low energy model Hamiltonian to the \emph{ab-inito} calculations of the bands. As usual, $v_{F}$  is the Fermi velocity. We find that $m^{\ast}c^{2}\approx 0.3$ MeV, that is, it is of the order of the bare electron mass ($m_0c^2\approx 0.5$ MeV; $m^\ast/m_0\approx0.6$), for $c$  the speed of light in vacuum. Defining an effective electron mass for the charge carriers in hBN is of paramount importance in the development of the method for addressing the excitonic boundstates in hBN that we will present below. The system we have in mind studying
is depicted in Fig. \ref{fig:Representation-of-the}. The top image represents  a hBN monolayer  deposited on a quartz substrate and illuminated by electromagnetic radiation. In the bottom one, the Brillouin zone of hBN is represented together with the electronic dispersion at the edges of this zone. The electromagnetic radiation can induce electronic transitions from the valence to the conduction band of the 2D crystal thus leading to the formation of excitons.

The paper is organized as follows: In Sec.\ref{sec:Formalism}, we develop the needed formalism for the calculation of the excitonic response of hBN on quartz. In Sec.\ref{sec:Results}, we provide results
for the absorption of hBN as function of frequency and show that the material can support both TE and TM exciton-polaritons of large wave number and, therefore, with a high degree of confinement. In Sec.\ref{sec:The-spontaneous-radiative}, we compute the spontaneous emission decay rate of excitons in hBN on quartz, a quantity essential for the interpretation of photoluminescent experiments. Finally, in the Appendix, we present some useful formulae used in the numerical part of our analysis and benchmark the method against known analytical results.

\section{Formalism\label{sec:Formalism}}

In this section we introduce the electronic model Hamiltonian and present the formalism for solving the 2D Wannier equation with an attractive electron-hole interaction described by the Rytova-Keldysh potential \cite{Rytova1967,Keldysh1979}. We also show how this Wannier equation can be obtained from the Bethe-Salpeter equation by means of a Fourier transform.

\subsection{Model Hamiltonian for hBN}

The electronic and optical properties of electrons in the hybridized $p_{z}$ orbitals of hBN are well described by a first-neighbor tight-binding model. This can be further simplified by writing a low energy model around the Dirac points, located at the corners of the Brillouin zone. This low energy model is, in this case, a massive Dirac equation in 2D, with an Hamiltonian given by:

\begin{equation}
H_{0}=v_{F}\hbar\mathbf{k}\cdot\bm{\sigma}+M\sigma_{z}\,,\label{eq:H0}
\end{equation}
for $v_{F}$, the Fermi velocity; $\mathbf{k}=(k_{x},k_{y})$ is the in-plane wave vector of the electrons in the hybridized  $p_{z}$ orbitals; $\bm{\sigma}=(\sigma_{x},\sigma_{y})$, with $\sigma_{j}$ the $j=x,y,z$ Pauli matrix: and $2M$ is the \emph{density functional theory }(DFT) band gap (e. g., at the LDA level of approximation) at the corners of the Brillouin zone. We note that the Pauli matrices do not refer to real spin but rather to the basis of the hexagonal lattice unit cell. The eigenproblem $H_0\vert u_{\mathbf{k},s}\rangle=E_{s}\vert u_{\mathbf{k},s}\rangle$, with $s=\pm,$ has a simple solution, reading 
\begin{eqnarray}
\vert u_{\mathbf{k},+}\rangle &=&\frac{1}{\sqrt{2E_{+}}}\left(\begin{array}{c}
\sqrt{M+E_{+}}\\
\frac{v_{F}\hbar ke^{i\theta}}{\sqrt{M+E_{+}}}
\end{array}\right)
\nonumber\\
&&\underset{k\rightarrow0}{\longrightarrow}\frac{1}{\sqrt{2M}}\left(\begin{array}{c}
\sqrt{2M}\\
\frac{v_{F}\hbar ke^{i\theta}}{\sqrt{2M}}
\end{array}\right)\,,\label{eq:spinor_plus}
\end{eqnarray}
and
\begin{eqnarray}
\vert u_{\mathbf{k},-}\rangle &=&\frac{1}{\sqrt{2E_{+}}}\left(\begin{array}{c}
-\frac{v_{F}\hbar ke^{-i\theta}}{\sqrt{M+E_{+}}}\\
\sqrt{M+E_{+}}
\end{array}\right)
\nonumber\\
&&\underset{k\rightarrow0}{\longrightarrow}\frac{1}{\sqrt{2M}}\left(\begin{array}{c}
-\frac{v_{F}\hbar ke^{-i\theta}}{\sqrt{2M}}\\
\sqrt{2M}
\end{array}\right)\,,\label{eq:spinor_minus}
\end{eqnarray}
with $E_{s}=s\sqrt{M^{2}+v_{F}^{2}\hbar^{2}k^{2}}$ (at spaces we also use $s=c/s$ when used as an index).

The asymptotic expression of the spinors when $k\rightarrow0$ will be used in the calculation of the radiative lifetime of the exciton. We note that this approximation is justified by the fact that the energy gap in hBN is rather large $\sim 4$ eV (in the absence of exchange corrections. These corrections further widen the gap).

The total electronic Hamiltonian is the sum of $H_{0}$ with the Rytova-Keldysh
interacting potential: 
\begin{equation}
V(r)=\frac{e^{2}}{4\pi\epsilon_{0}}\frac{\pi}{2}\frac{1}{r_{0}}\left[H_{0}\left(\frac{\epsilon_{1}r}{r_{0}}\right)-Y_{0}\left(\frac{\epsilon_{1}r}{r_{0}}\right)\right]\,,\label{eq:RK}
\end{equation}
where $r_{0}$ is a length scale of the order of $r_{0}\sim\epsilon_{2}d/2$,
with $d$ the thickness of the 2D material, and $\epsilon_{2}$ the
dielectric function of the 2D material film; $\epsilon_{1}$ is the
dielectric function of the medium surrounding the 2D material; $H_{0}(x)$
is the Struve function and $Y_{0}(x)$ is the Bessel function of the
second kind. This form of the potential follows from the solution
of Laplace equation for a thin film embed in a medium. If the 2D material
is encapsulated between two different dielectrics, $\epsilon_{1}$
and $\epsilon_{3}$, then it is necessary to replace $\epsilon_{1}$
in Eq. \ref{eq:RK} by the arithmetic mean of the two dielectric functions.

\subsection{Derivation of the Wannier equation in 2D}

In second quantization, the state of an exciton of momentum $\mathbf{Q}$, in motion in an hBN layer of area $A$, can be expressed by 
\begin{equation}
\vert\nu,\mathbf{Q}\rangle=\frac{1}{\sqrt{A}}\sum_{\mathbf{k}}\phi_{\nu}(\mathbf{k})a_{\mathbf{k}+\mathbf{Q},c}^{\dagger}a_{\mathbf{k},v}\vert GS\rangle,\label{eq:exciton}
\end{equation}
where the state $\vert GS\rangle$ represents the electronic ground state of hBN, that is, a fully occupied valence band and an empty conduction band. Also, $\phi_{\nu}(\mathbf{k})$ is the Fourier transform
of the real space exciton wave function characterized by the set of quantum numbers $\nu$ (in our case it will be the principal and magnetic quantum numbers). The second quantized operators $a_{\mathbf{k}+\mathbf{Q},c}^{\dagger}$ and $a_{\mathbf{k},v}$ create and annihilate an electron of momentum $\mathbf{k+\mathbf{Q}}$ in the conduction band and an electron of
momentum $\mathbf{k}$ in the valence band, respectively. This state can then be expressed as $\vert\nu,\mathbf{Q}\rangle=b_{\mathbf{Q},\nu}^{\dagger}\vert GS\rangle,$
for \textbf{$b_{\mathbf{Q},\nu}^{\dagger}$ } the a bosonic operator,

\begin{equation}
b_{\mathbf{Q},\nu}^{\dagger}=\frac{1}{\sqrt{A}}\sum_{\mathbf{k}}\phi_{\nu}(\mathbf{k})a_{\mathbf{k}+\mathbf{Q},c}^{\dagger}a_{\mathbf{k},v}.\label{eq:boson}
\end{equation}
This representation will be useful in the calculation of the radiative lifetime of the exciton. We also note that the bosonic nature of the operator (\ref{eq:boson}) is guaranteed only in an average over the ground state.

The Hamiltonian describing the electrons in hBN can then written in second quantization as 
\begin{equation}
H=H_{0}+V\,,
\end{equation}
 with
\begin{equation}
H_{0}=\sum_{\lambda,\mathbf{k}}E_{\lambda,\mathbf{k}}\hat{a}_{\lambda,\mathbf{k}}^{\dagger}\hat{a}_{\lambda,\mathbf{k}}\,,
\end{equation}
where $\lambda=c,v$, and $E_{\lambda,\mathbf{k}}=E_{s}$, and with
the interaction term given by
\begin{eqnarray}
V  &=&\frac{1}{2A}\sum_{\mathbf{k}_{1},\mathbf{k}_{2},\mathbf{p}}\sum_{\lambda_{1}\lambda_{2}\lambda_{3}\lambda_{4}}V(\mathbf{p})F_{\lambda_{1}\lambda_{2}\lambda_{3}\lambda_{4}}(\mathbf{k}_{1},\mathbf{k}_{2},\mathbf{p})
\nonumber\\
&\times&\hat{a}_{\mathbf{k}_{1}+\mathbf{p},\lambda_{1}}^{\dagger}\hat{a}_{\mathbf{k}_{2}-\mathbf{p},\lambda_{2}}^{\dagger}\hat{a}_{\mathbf{k}_{2},\lambda_{3}}\hat{a}_{\mathbf{k}_{1},\lambda_{4}}\,,
\end{eqnarray}
and 
\begin{equation}
F_{\lambda_{1}\lambda_{2}\lambda_{3}\lambda_{4}}(\mathbf{k}_{1},\mathbf{k}_{2},\mathbf{p})=u_{\mathbf{k}_{1}+\mathbf{p},\lambda_{1}}^{\dagger}u_{\mathbf{k}_{2}-\mathbf{p},\lambda_{2}}^{\dagger}u_{\mathbf{k}_{2},\lambda_{3}}u_{\mathbf{k}_{1},\lambda_{4}}\,,
\end{equation}
is a product of spinors (\ref{eq:spinor_plus}) and (\ref{eq:spinor_minus}) and is termed the form factor. The function $V(\mathbf{p})$ is the Fourier transform of the Rytova-Keldysh potential (\ref{eq:RK}) is known analytically \cite{Rytova1967}, and it is given by $V(\mathbf q)=e^2/[2\epsilon_0q(r_0q+\epsilon_1)]$. Next we show that if the state (\ref{eq:exciton}) is an eigenstate of $H$, then $\phi_{\nu}(\mathbf{k})$ obeys the Bethe-Salpeter equation
and its Fourier transform to real space gives rise to the Wannier equation.

In order to obtain the Bethe-Salpeter equation we proceed as follows: we assume that the state (\ref{eq:exciton}) is an eigenstate of the
Hamiltonian $H$. If this is the case, then $H$ can be written as $H=\sum_{\mathbf{Q},\nu}E_{\mathbf{Q},\nu}b_{\mathbf{Q},\nu}^{\dagger}b_{\mathbf{Q},\nu}$, where $E_{\mathbf{Q},\nu}$ are the energy eigenvalues of the exciton. Next we compute the commutator of $H$ with $b_{\mathbf{Q},\nu}^{\dagger}$ using both the fermionic and bosonic representations. In the end we must demand both results  be the same. Proceeding as such, we obtain the following equation for the wave function of the exciton in momentum space:
\begin{eqnarray}
E\phi_{\nu}(\mathbf{k})=\phi_{\nu}(\mathbf{k})(E_{c,\mathbf{k}}-E_{v,\mathbf{k}})+\frac{1}{A}\phi_{\nu}(\mathbf{k})\sum_{\mathbf{p}}V(\mathbf{p})\times
\nonumber\\
\left[u_{\mathbf{k},v}^{\dagger}u_{\mathbf{k-p},v}^{\dagger}u_{\mathbf{k},v}u_{\mathbf{k-p},v}-u_{\mathbf{k},c}^{\dagger}u_{\mathbf{k+p},v}^{\dagger}u_{\mathbf{k},c}u_{\mathbf{k+p},v}\right]
%\times
\nonumber \\
-\frac{1}{A}\sum_{\mathbf{p}}V(\mathbf{p})\phi_{\nu}(\mathbf{p+k})u_{\mathbf{p+k},v}^{\dagger}u_{\mathbf{k},c}^{\dagger}u_{\mathbf{p+k},c}u_{\mathbf{k},v}\,,
\end{eqnarray}
where $E$ represents the exciton energy eigenvalues. The latter equation --- an eigenvalue equation in momentum space for $\phi_{\nu}(\mathbf{k})$ --- is known as the Bethe-Salpeter equation. The physics of the different terms is clear: the first term is the energy when a particle-hole excitation is created, in the non-interacting limit. The second term represents the exchange energy correction to the non-interacting particle-hole excitation energy (its value determines the magnitude of the gap), the third and final term represents the attraction between the electron created in the conduction band and the hole left behind in the valence band. Crucially this term in negative, although in original Hamiltonian the interaction between electrons is obviously repulsive. Note that, although this problem can be solved directly in the momentum space, it involves an integral equation for $\phi_{\nu}(\mathbf{k})$, whose solution can be rather delicate \cite{Andre2017}, as the sum over the wave vector ranges over an infinite area in the reciprocal space. Another approach is to transform this integral equation into a differential one  going to real space. Now, Fourier transforming this equation is also no simple task (and virtually an impossible one) due to the presence of the spinors. To do so, we make the following observation concerning the form factors in the Bethe-Salpeter equation. In the case of a large energy gap, one can take,

\begin{equation}
u_{\mathbf{p+k},v}^{\dagger}u_{\mathbf{k},c}^{\dagger}u_{\mathbf{p+k},c}u_{\mathbf{k},v}\longrightarrow1+\mathcal{O}\left(1/M^{2}\right)\,.
\end{equation}
so as to forego the spinorial structure of the last term of the Bethe-Salpeter equation. It is also considered that both the energy difference, $E_{c,\mathbf{k}}-E_{v,\mathbf{k}}$, and the exchange energy corrections are expanded up to second order in $\mathbf{k}$.
The resulting differential equation --- in real space --- reads as,

\begin{equation}
(E-E_{g})\psi_{\nu}(\mathbf{r})=-\bigg[\frac{\hbar^{2}}{2\mu}\nabla^{2}+V(\mathbf{r})\bigg]\psi_{\nu}(\mathbf{r})\,,
\end{equation}
which is known as the Wannier equation for the excitonic wave function. Here $\mu$ is the reduced mass of the exciton which, in our  model, reads $m^{\ast}/2$. The quantity $E_{g}$ (termed \emph{free particle band gap}) reads $E_{g}=2M+\Delta_{ex}$, for $\Delta_{ex}$, the exchange energy correction to the DFT gap evaluated at $\mathbf{k}=0$. This correction is given by, 

\begin{eqnarray}
\Delta_{ex}&=&\frac{\alpha M}{r_{0}}\frac{c}{v_{F}}\frac{{\rm arcsinh}\frac{\epsilon_{1}v_{F}\hbar}{r_{0}M}+{\rm arcsinh}\frac{r_{0}M}{\epsilon_{1}v_{F}\hbar}}{\sqrt{\Big(\frac{\epsilon_{1}}{r_{0}}\Big)^{2}+\Big(\frac{M}{v_{F}\hbar}\Big)^{2}}}
\nonumber\\
&\approx &\frac{\alpha\hbar c}{r_{0}}\log\frac{2Mr_{0}}{\epsilon_{1}v_{F}\hbar}+\frac{\alpha\epsilon_{1}\hbar^{2}v_{F}c}{r_{0}^{2}M}\,,\label{eq:Exchange_gap}
\end{eqnarray}
where $\alpha$ is the fine structure constant.
The approximate result holds for $Mr_{0}\gg\epsilon_{1}\hbar v_{F}$, for $\epsilon_{1}$, the average value of the dielectric functions of the substrate and capping layer.

In the next section we describe a semi-analytical method of solving the Wannier equation, whose solutions will allow us to easily access the optical properties of 2D materials.

\subsection{An efficient method for solving the Wannier equation in 2D}

We have shown in the previous section that solving the excitonic problem requires the determination of the exciton wave function in both real and reciprocal spaces. As such, it would be convenient that the wave function could be expressed in a form that would ease the calculations, as a fully numerical approach usually renders the calculation numerically expensive. Here, we show that a quasi-analytical expression for the wave functions of the exciton can be written using a set of Gaussian functions. The initial idea is relatively simple: in solid state physics, DFT calculations of the band structure of solids are routinely performed in different basis sets --- plane-waves, Slater-type orbitals, and Gaussian functions being the most commonly used. Plane waves are clearly not suitable for our problem, as we are dealing here with bound-states. The Slater basis has the advantage of describing the behavior of the wave functions near the origin in a very precise manner, but it is computationally demanding when the matrix elements of the kinetic operator have to be determined. The basis of Gaussian functions requires considerably more terms to describe the behavior of the wave function near the origin, but has the advantage of allowing for analytical expressions for the matrix elements of both
the kinetic energy operator and the electron-electron interacting
potential. This latter choice of basis is the approach used in this
this paper. Drawing inspiration from the solution of the 2D hydrogen
atom (see Appendix), we write our wave function in the form:
\begin{equation}
\psi_{\nu}(\mathbf{r})=\mathcal{A}_{\nu}\sum_{j=1}^{N}c_{j}^{\nu}e^{im\theta}r^{|m|}e^{-\zeta_{j}r^{2}}\,,
\end{equation}
Note that $e^{im\theta}r^{|m|}$ follows from the eigenfunctions
of the $z-$component of the angular momentum and describes the wave function near the origin, for $m=0,\pm1,\pm2,\ldots$, the
magnetic quantum number; the Gaussian term $e^{-\zeta_{j}r^{2}}$
describes the decay of the wave function at large distances, with
with a decay constant that depends on $\zeta_{j}$. The
coefficients $c_{j}^{\nu}$ are weights for the different Gaussians;
whereas ${\cal A}_{\nu}$ is a normalization constant. An additional
advantage of this method is that the matrix elements of both the kinetic
operator and the electron-electron interaction do not mix different
$m$ values and therefore, the eigenvalue problem is block diagonal in
the angular momentum space. The quantum number $\nu$ plays the role
of the principal quantum number: for each value of $m$ there is an
infinite number of $\nu$'s (note, however, that we also use the label
$\nu$ for denoting both the magnetic and the principal quantum numbers
together).

\begin{figure*}
\includegraphics[scale=0.7]{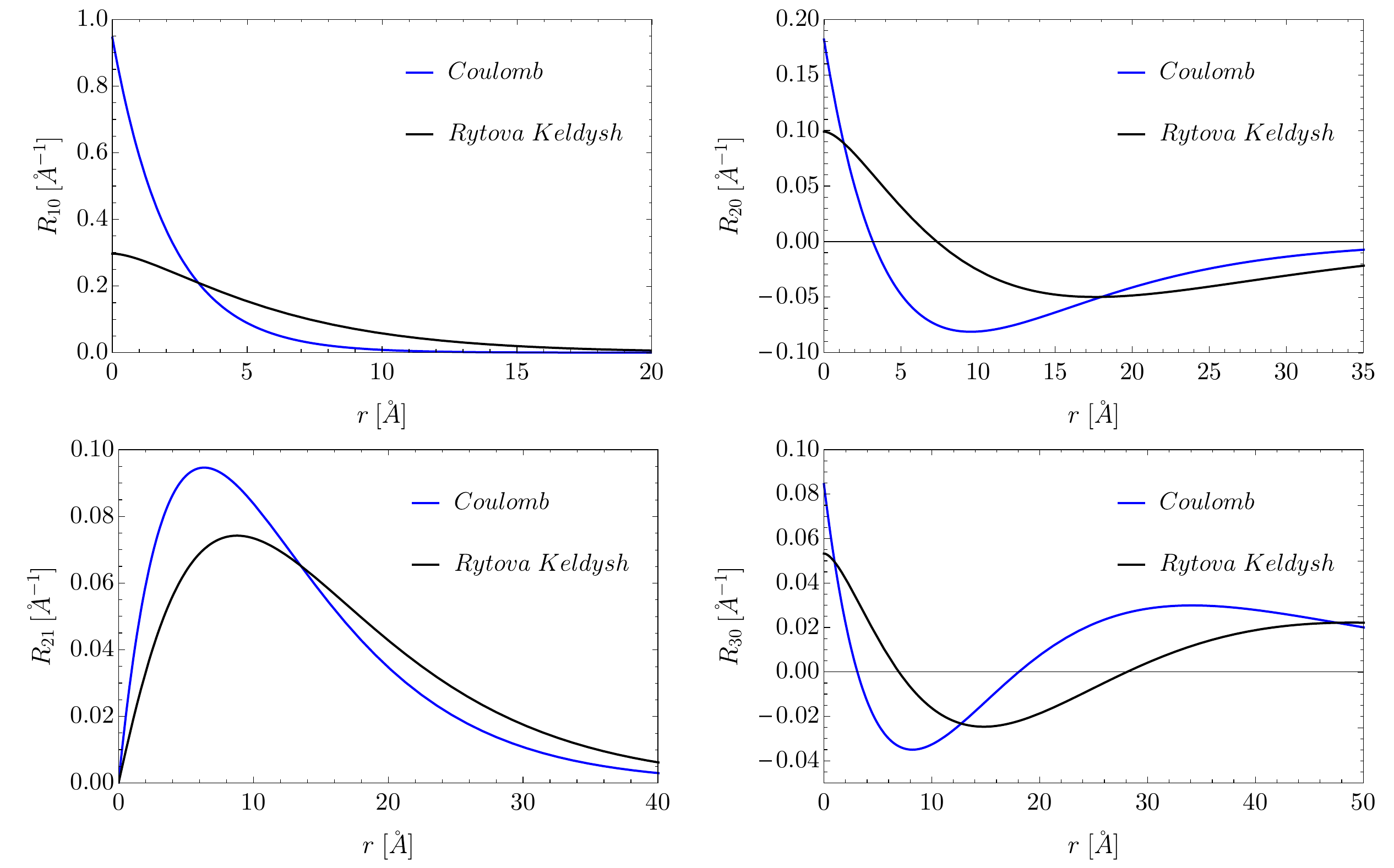}

\caption{Few radial wave functions of the Rytova-Keldysh and Coulomb potentials
for the parameters of Table \ref{tab:Parameters-used-in}. 
With the given set of parameters the wave functions of the Rytova-Keldysh potential are more spread out in space, which imply larger
(less bound) energy values (compare values of Table \ref{tab:Energy-spectrum-RK} and Table \ref{tab:AnalyticalVsNumerical Energy Spectrum}).
The calculation used an effective mass $m^\ast=0.6m_0$.
\label{fig:Few-radial-wave-Rytova}}

\end{figure*}

Using our trial wave function and computing the matrix elements of
the kinetic and potential energy operators, the generalized eigenvalue
problem acquires the form
\begin{equation}
\sum_{j=1}^{N}[H(\zeta_{i},\zeta_{j})-S(\zeta_{i},\zeta_{j})E]c_{j}^{\nu}=0,\label{eq:Wheeler}
\end{equation}
where $H(\zeta_{i},\zeta_{j})$ is called the Hamiltonian kernel and
$S(\zeta_{i},\zeta_{j})$ is the superposition kernel. This differs
from a Kronecker$-\delta$ kernel since the set of Gaussian functions is not an orthogonal
basis. Both kernels have an analytical expression given in the Appendix.
Equation (\ref{eq:Wheeler}) has first been written in nuclear physics
and is termed the Griffin-Hill-Wheeler equation \cite{Griffin1957}. The normalization
constant is determined in the usual way and reads

\begin{equation}
\mathcal{A}_{\nu}=\sqrt{\frac{1}{\pi\:\mathcal{S}_{\nu}}}\,,
\end{equation}
with $\mathcal{S}_{\nu}=\sum_{j=1}^{\infty}\sum_{j'=1}^{\infty}c_{j}^{\nu*}c_{j'}^{\nu}(\zeta_{j}+\zeta_{j'})^{-1-|m|}\Gamma(|m|+1)$,
and $\Gamma(x)$ the gamma-function. A critical aspect of the method
is the judicious choice of the parameters $\zeta_{j}$. A choice not
so well known is the use of a logarithmic grid of $\zeta's$ according
to the rule \cite{Mohallem1986}
\begin{equation}
\Omega=\frac{\ln\zeta}{A},\qquad A>1\,,
\end{equation}
where the $\Omega's$ are uniformly distributed in an interval $[\Omega_{{\rm min}},\Omega_{{\rm max}}]$
and $A$ is typically chosen in the interval $[6,8].$ With the method
exposed, we move on to the calculation of the eigenvalues and eigenvectors
of the Griffin-Hill-Wheeler equation for both the Coulomb (see Appendix) and the Rytova-Keldysh
potentials. In Table \ref{tab:Energy-spectrum-RK}, we present the spectrum
of the Rytova-Keldysh potential in the  conditions discussed
in Sec. \ref{sec:Results}. In Fig. \ref{fig:Few-radial-wave-Rytova}, we
depict a set of solutions of Eq. (\ref{eq:Wheeler}) for both the Rytova-Keldysh and Coulomb potentials, where again used the conditions of the following section. In Table \ref{tab:Energy-spectrum-RK} we present the bound state energies for different reduced masses $\mu$: as $\mu $ decreases the kinetic energy increases and the states become less bound.

\begin{table*}
\centering{}%
\begin{tabular}{|c|c|c|c|c|c|}
\hline
$m^\ast$&principal q.n. & $n=1$ & $n=2$ & $n=3$ & $n=4$\\
\hline
$m_0$&Energy (eV) & -0.992 & -0.274 & -0.126 & -0.072\\
\hline
$0.7m_0$&Energy (eV) &  -0.845&  -0.213 & -0.094  & -0.052\\
\hline
$0.5m_0$&Energy (eV) & -0.718 &-0.155 & -0.071 & -0.0314\\
\hline
\end{tabular}
\caption{\label{tab:Energy-spectrum-RK}Energy spectrum of the Rytova-Keldysh
potential for hBN on quartz, with $r_{0}=10$ \AA\, and with the average dielectric function of substrate and capping layer taken to
be $\epsilon_{1}=2.4$, that is, the substrate is quartz and the capping
layer is air. Different values of the effective mass $m^\ast$ have been considered.
The calculation used $100$ Gaussians and the first four energy eigenvalues labeled by the princial quantum number (q.n.) $n$ are given (they all correspond to the $m=0$ case).}
\end{table*}

\section{Results\label{sec:Results}}

In this section, we compute the absorption of hBN on quartz. We also discuss the existence
of surface exciton-polaritons and their application in sensors.

\subsection{Absorption of hBN: Theory}

\begin{figure}
\includegraphics[scale=0.6]{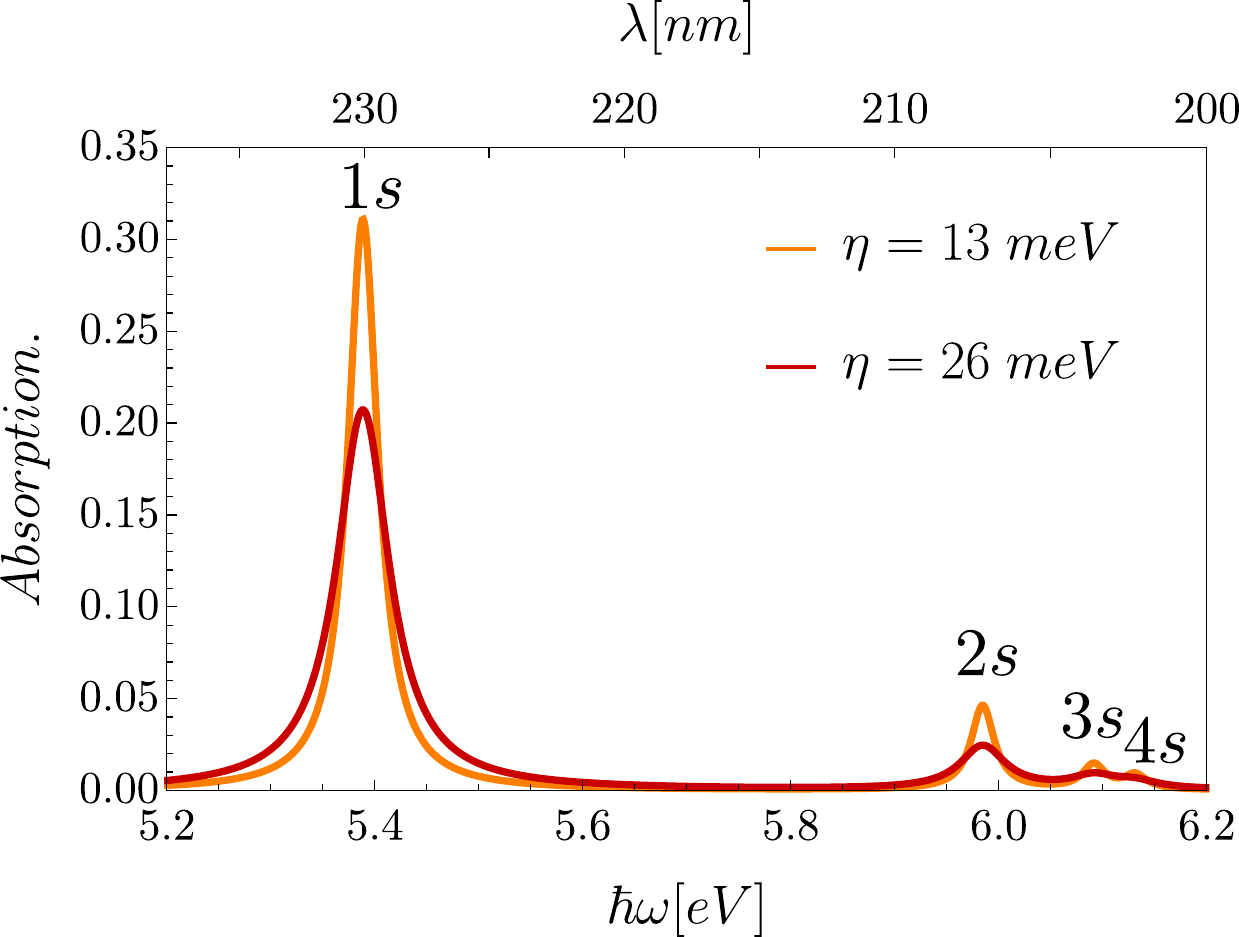}

\caption{Absorption of hBN on quartz. \label{fig:Absorption-of-hBN} In principle
there is only one fitting parameter, the non-radiative decay rate
$\eta$, which was chosen as $\eta=13$ meV. The value of the free particle band gap is given by $E_{g}=2M+\Delta_{ex}$,
which has been computed to be $6.2$ eV, for hBN on quartz, and $7.2$
eV for hBN in vacuum, using the parameters of Table \ref{tab:Parameters-used-in}.  An effective mass of $m^\ast=0.6m_0$ was used, where $m_0$ is the bare electron mass. The effect of changing the reduced mass is discussed in Table \ref{tab:Energy-spectrum-RK}. The Rydberg series of the excitonic spectrum is clearly seen.}
\end{figure}

The calculation of the absorption of hBN requires the calculation
of the reflection, ${\cal R}$, and transmission, ${\cal T}$, coefficients,
with the absorption defined as

\begin{equation}
{\cal A}=1-{\cal R}-\Re(\sqrt{\epsilon_3/\epsilon_1}){\cal T}.\label{eq:Abs}
\end{equation}
The determination of ${\cal R}$ and ${\cal T}$ requires the solution
of the Fresnel problem (that is, the scattering of electromagnetic
radiation by a flat interface) when a 2D material is embed in two
different dielectrics. From the solution of the Fresnel problem at
normal incidence the absorption is given by Eq. (\ref{eq:Abs}) with
${\cal R}=\vert r\vert^{2}$ and
\begin{equation}
r=\frac{\sqrt{\epsilon_{3}}-\sqrt{\epsilon_{1}}+\sigma/(\epsilon_{0}c)}{\sqrt{\epsilon_{3}}+\sqrt{\epsilon_{1}}+\sigma/(\epsilon_{0}c)}\,,
\end{equation}
and ${\cal T}=\vert t\vert^{2}$, with

\begin{equation}
t=\frac{2\sqrt{\epsilon_{1}}}{\sqrt{\epsilon_{3}}+\sqrt{\epsilon_{1}}+\sigma/(\epsilon_{0}c)}\,,
\end{equation}
with $\epsilon_1$ and $\epsilon_3$ the dielectric functions of air and quartz, respectively, and $\sigma=\sigma(\omega)$ is the optical conductivity of the 2D
material, which can be obtained by usual perturbation theory. After a lengthy calculation we obtain,
\begin{eqnarray}
\frac{\sigma}{\sigma_{0}}=\frac{i}{4\pi^{3}}\sum_{\nu}|E_{g}+E_{\nu}|\Lambda_{\nu}\left(\frac{1}{\hbar\omega+E_{\nu}+E_{g}+i\eta}+
\right.
\nonumber\\
\left.
\frac{1}{\hbar\omega-E_{\nu}-E_{g}+i\eta}\right)\,,\label{eq:sigma_exciton}
\end{eqnarray}
where $\Lambda_{\nu}$  is roughly proportional to the square modulus of the  wave function
of the exciton; $E_{g}$ is the gap between the valence and the conduction bands (often called the free particle band gap to distinguish it from
the the optical band gap, the latter determined by excitons) after
the inclusion of the exchange energy correction; $E_{\nu}$ is the exciton energy
level associated with the quantum number $\nu$ (including both the
principal and magnetic quantum numbers).  Only three magnetic quantum numbers produce a non-zero result, $m=0$ and $m=\pm2$, the largest contribution being, by far, that of $m=0$; finally, $\eta$ is the non-radiative
decay rate, encompassing all possible decay channels. The conductivity here
is expressed in units of graphene universal conductivity $\sigma_{0}=e^{2}/(4\hbar)$.
Using Eq. (\ref{eq:sigma_exciton}) in Eq. (\ref{eq:Abs}) the absorption
of hBN can be computed. This is given in Fig \ref{fig:Absorption-of-hBN}. The parameters used in the theoretical calculations
are given in Table \ref{tab:Parameters-used-in}.  As for the
non-radiative decay rate, $\eta$, it is a fitting parameter and whose microscopic
calculation is beyond the present study, as it requires calculations
taking into account electron-impurity, electron-phonon and electron-electron
scattering processes. In practical terms, the value of $\eta$ controls
the intensity of the absorption curve. For the dielectric function
of quartz in the ultraviolet, we have used the experimentally tabulated values.
We note that the electron-electron interaction depends on the static
dielectric function of quartz, $\epsilon_{{\rm quartz}}$, whereas
the absorption depends on the dynamic dielectric function, $\epsilon(\omega)$.

\begin{table*}
\centering{}%
\begin{tabular}{|c|c|c|c|c|c|c|c|c|c|}
\hline 
$\mathbf{N}$ & $\mathbf{\Omega_{min}}$ & $\mathbf{\Omega_{max}}$ & $\mathbf{A}$ & $\mathbf{E_{g}}$ (eV) & $\mathbf{M}$ (eV) & $\mathbf{\hbar v_{F}}$ (eV$\cdot$\AA) & $\boldsymbol{\eta}$ (eV) & $\mathbf{r}_{0}$ (\AA) & $\epsilon_{{\rm quartz}}$\tabularnewline
\hline 
100 & -2 & 5 & 6 & 6.17 & 1.96 & 5.06 & 0.013 & 10 & 3.9\tabularnewline
\hline 
\end{tabular}
\caption{Parameters used in the calculation of the excitonic conductivity of
hBN. The parameters $r_{0}$, $M$, and $\hbar v_{F}$ where extracted
from Ref. \cite{Ferreira2019}. The parameter $\epsilon_{{\rm quartz}}$
refers to the static dielectric constant of quartz that enters in
the Rytova-Keldysh potential. The parameters $N$, $\Omega_{{\rm min}}$,
$\Omega_{{\rm max}}$, and $A$ were chosen with the criteria of reproducing
accurately the analytical 2D energy spectrum and wave functions of
the attractive Coulomb problem. We have varied these set of parameters
and checked that significant variations do not alter the final result.
The non-radiative decay rate $\eta$ is a fitting parameter and its
microscopic calculation is beyond the scope of the present study.
\label{tab:Parameters-used-in}}
\end{table*}

As given above,  we have determined an analytical expression for the exchange energy
correction to the DFT band gap from Eq. (\ref{eq:Exchange_gap}) and
have found that the new gap, $E_{g}=2M+\Delta_{ex}$, reads $\sim6.2$
eV for hBN on quartz. For hBN in vacuum, we found that
$E_{g}=2M+\Delta_{ex}\sim7.2$ eV. This value is in agreement with
DFT+GW calculations, as can be seen in Table \ref{tab:Energy-gap-parameter}. We note in passing that a calculation of the exchange energy was made
in Ref. \cite{Andre2017} in the context of the absorption of electromagnetic
radiation by TMD and the resulting energy gap $E_{g}$ was shown to
be in excellent agreement with the experimental data.

\begin{table}
\centering{}%
\begin{tabular}{|c|c|c|c|c|c|c|}
\hline 
Ref. & \cite{Winther2017} & \cite{Galvani2016} & \cite{Mengle2019} & \cite{Weston2018}& \cite{Guilhon2019}&this work\tabularnewline
\hline 
\hline 
 $E_{g}$ (eV) & 7-7.5 & 7.25 & 8($\dagger$) & 6.46& 7.4 &7.2,8.0$(\ast)$\tabularnewline
\hline 
\end{tabular}
\caption{Theoretical energy gap parameter, $E_{g}$, computed in different references.
($\ast$) Values in vacuum for $r_0=10,6.9\,$\AA, respectively (see also Table \ref{tab:Energy-levels-on-HOPG}).  ($\dagger$) Direct gap; these authors found an indirect band gap of 7.74 eV for the monolayer.
(The value of $E_g$ on quartz was found by us to be $E_g\approx6.2$ eV.)
\label{tab:Energy-gap-parameter}}

\end{table}

\subsection{Photoluminescence of hBN on graphite}

For putting our theory to the test, we have compared the
theoretical  position of the excitonic energy levels  of a hBN single-layer to  photoluminescence (PL)  of  hBN single layer  on highly oriented pyrolytic graphite (HOPG),
published very recently \cite{Elias2019}. In this paper, the authors were able to measure the PL of a single hBN layer in the frequency range
$\lambda\in[240,200]$ nm, at a temperature of 10 K.  They identified a broad deep-ultraviolet
emission in the region $\lambda\in[240,210]$ nm. This type of emission is, traditionally, 
attributed to to excitons trapped in defects 
of the hBN lattice  \cite{Watanabe2004}, energetically located inside the gap, and is termed, in bulk hBN, the $D-$series ($D$ from diffuse)
\cite{Shue2016}. In addition
to this,  those authors identified two sharp  peaks ($S-$series) in the region 
$\lambda\in[210,200]$ nm. This emission is attributed to excitons.
To explain the data of Ref. \cite{Elias2019}, we have computed absorption (which depends directly on the position of the excitonic energy levels)  of a single hBN layer on HOPG, which can, to first approximation be considered as a perfect conductor. This fact leads to screening the 
Rytova-Keldysh potential, which has an approximate analytical solution for $d\ll r_0$:
\begin{eqnarray}
V(\mathbf r)&=&\frac{e^2}{\epsilon_0} \int_0^\infty \frac{dq}{2\pi} J_0(qr) \left[ 2r_0 q + \frac{e^{qd}}{\sinh(qd)} \right]^{-1}\nonumber\\
&\approx & \frac{e^2}{4\pi \epsilon_0 r_0}K_0(r/\sqrt{2dr_0})\,,
\label{eq:RK_screened}
\end{eqnarray}
with $d$ the distance between hBN and HOPG substrate and $K_0(x)$ the modified Bessel function.
We have verified numerically that for hBN at a distance of $d=3.5$\AA\,  
 of the HOPG substrate (this distance corresponds to the measured step height (3.5\AA) between HOPG  and hBN \cite{Elias2019,Li2016} using AFM), the screened potential is well described by the approximation in Eq. (\ref{eq:RK_screened}), although slightly more attractive. In addition, for the approximated screened potential, the exchange energy also has an analytical form:
 \begin{equation}
 \Delta_{ex}^{scr}=-i\frac{\sqrt{2d}M\alpha \hbar c\,
{\rm arccos}\frac{M\sqrt{2dr_0}}{\hbar v_F}
 }{\sqrt{r_0}\sqrt{2dr_0M^2-\hbar^2v_F^2}}\,, 
 \end{equation}
 with the corrected gap given by $E_g^{src}=2M+
 \Delta_{ex}^{src}$ and the constraint $2dr_0M^2>\hbar^2v_F^2$;
 the Fourier transform of $V(\mathbf r)$ given by Eq. (\ref{eq:RK_screened}) reads $V(\mathbf q)=4 \pi \hbar c \alpha (1/q) [2r_0 q + 1/(qd)]^{-1}$.
\begin{table}
\centering{}%
\begin{tabular}{|c|c|c|c|}
\hline 
 $n=1$  & $E_g^{scr}$ (eV) & $r_0\,$(\AA) & $d\,$(\AA)\\
\hline
 -1.546 eV   &7.6($\ast$)  &6.9($\dagger$) & 3.5\\
 \hline
\end{tabular}
\caption{\label{tab:Energy-levels-on-HOPG}
Excitonic energy levels of hBN at a distance of $3.5$\AA\, from
a highly conductive HOPG substrate (a value fixed by the experimental measured distance using AFM). The energy level
is depicted in Fig. \ref{fig:Absorption-PL} as peaks in the absorbance spectrum.  The effective mass of the electrons is $m^\ast=0.6m_0$. Note that we have changed $r_0$ 
relative to the value give in Table \ref{tab:Parameters-used-in}. 
This was necessary for making
the $1s-$level coincident with the excitonic peak in the PL data of Fig. \ref{fig:Absorption-PL}.  However, this 
choice of $r_0$ is consistent with the available data in the literature. Indeed, the value of $r_0$ can be estimated from \cite{Cudazzo2011}
$r_0= (\sqrt{\epsilon_\perp\epsilon_\parallel}-1) d/2=6.5$, with $\epsilon_\perp=3.29$ and $\epsilon_\parallel=6.82$ \cite{Laturia2018}. If, on the other hand, we use the results of
Refs. \cite{Pedersen2017,Latini2015}, we find $r_0=(\kappa-1) d/2=6.8$, with $\kappa=4.9$. Both results for $r_0$ are in approximate  agreement with our choice.
($\ast$) The gap in vacuum for $r_0=6.9$\AA\, reads $E_g=$7.97 eV, larger than in the presence of the HOPG, as it should be. We note that, incidentally,
this value of $E_g$ coincides with that of Refs. \cite{Mengle2019,Pedersen2017} (see Table \ref{tab:Energy-gap-parameter}). ($\dagger$) A  self-consistent calculation
using no approximations to the screened potential and with $r_0$ determined via 
a RPA calculation, gives $r_0=7.2$\AA\, and $E_g^{scr}\approx7.2$ eV \cite{Chaves2019}. 
}
\end{table}
\begin{figure}
\includegraphics[scale=0.6]{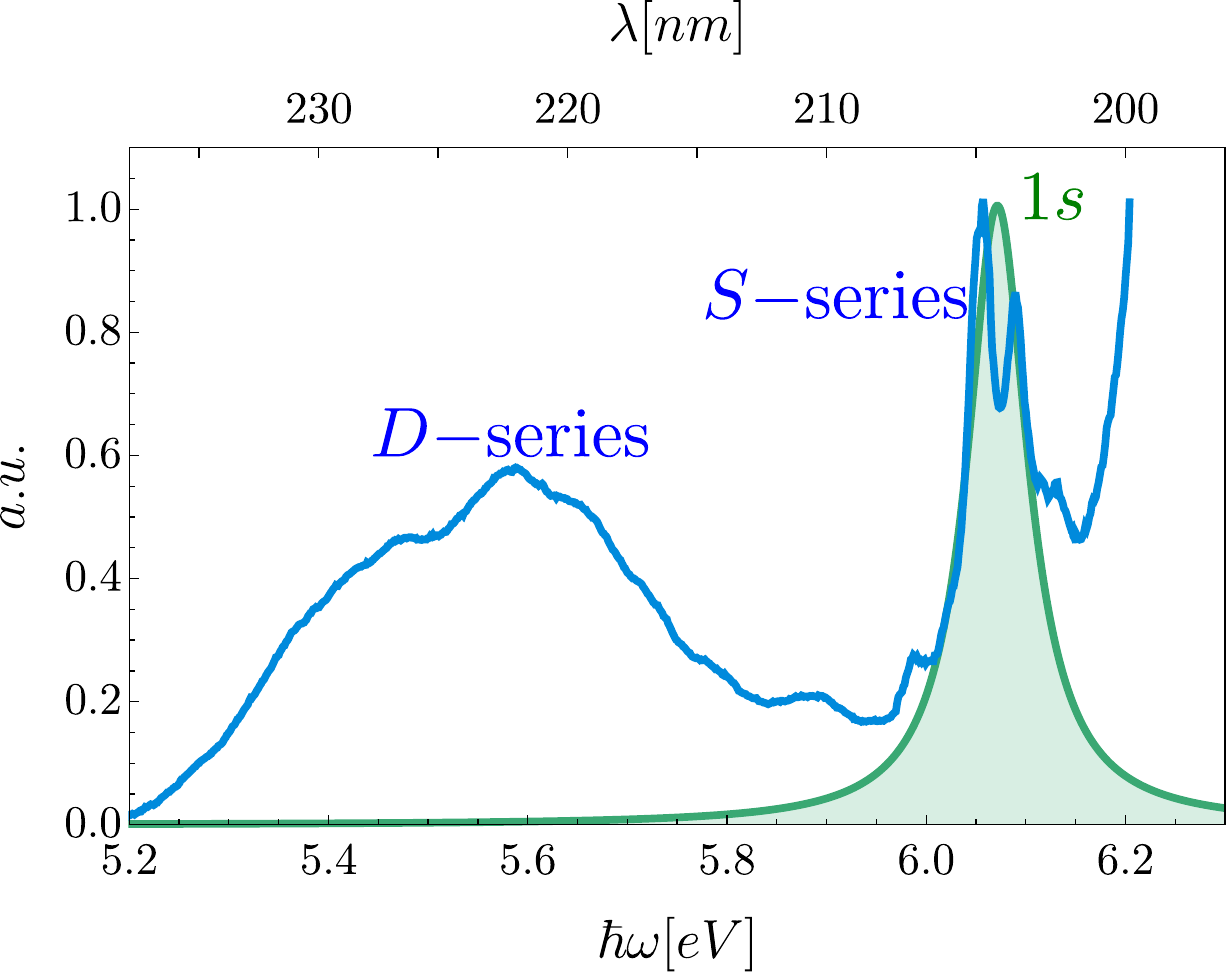}
\caption{Experimental photoluminescence (blue line) 
\cite{Elias2019}
and theoretical resonant absorption showing a peak at the energy level $n=1$ given in Table \ref{tab:Energy-levels-on-HOPG}.  An effective mass  $m^\ast=0.6m_0$ was used, in agreement with the estimations made in Sec. \ref{sec:Intro}. Both the absorption (theory) and the PL spectrum (experiment)
have been normalized to their maximum value. The computed gap
with exchange corrections reads $E_g^{src}=7.6$ eV.
A value of $\eta=30$ meV has been used, in agreement with the experimental measurements \cite{Elias2019} at 10 K. 
The diffuse series ($D-$series) is due to excitons trapped in defects
forming inside the gap; the sharp series ($S-$series) is due to (free) excitons.
\label{fig:Absorption-PL}}
\end{figure}
The result of our calculation is given in Table \ref{tab:Energy-levels-on-HOPG} and depicted in Fig. \ref{fig:Absorption-PL}
together with the PL data. Our calculation reveals a single  excitonic $s-$level given by the value of Table \ref{tab:Energy-levels-on-HOPG}. The $1s$ level coincides with the  $S-$series. For the approximated form of the interaction potential only the $1s$ bound state exists.
 Finally, we note that the double peak seen in the experiment is due to out-of-plane phonons coupling an electron in the valley $\mathbf{K}$ and a hole in the valley $\mathbf{K}'$, or vice-verse.   Our theory does not take phonons into account therefore is unable to describe
the peak splitting.
We stress that the apparent increase of PL near $200$ nm seen in the PL data is due to stray light from the excitation laser and is not an excitonic effect.

\subsection{Exciton-polaritons in hBN}

As discussed in the Introduction, some peptides have a strong absorption
at the frequency of the excitonic resonances of hBN on quartz.
When this situation happens a sensor working in the corresponding
spectral range is conceivable. A widely used type of sensor is based
on the concept of polariton, and it is associated to two of its fundamental properties: (i) an evanescent decay of
the electromagnetic field around the surface of the 2D material; (ii)
an enhancement of electromagnetic field in the vicinity of the 2D
material. These two properties should allow the determination of very small changes in
the refractive index of the medium  surrounding the 2D material and should, therefore, work as a sensor of small quantities of
 analyte.

There are different kinds of polaritons: phonon-polaritons, plasmon-polatirons,
magnon-polaritons, and exciton-polaritons. All of them have in common
the two properties mentioned above. In our case, we are concerned
with the exciton-polariton. We shall show below that the electromagnetic
field of the polariton is concentrated within few nanometers from the
surface of hBN.

\begin{figure}
\includegraphics[scale=0.6]{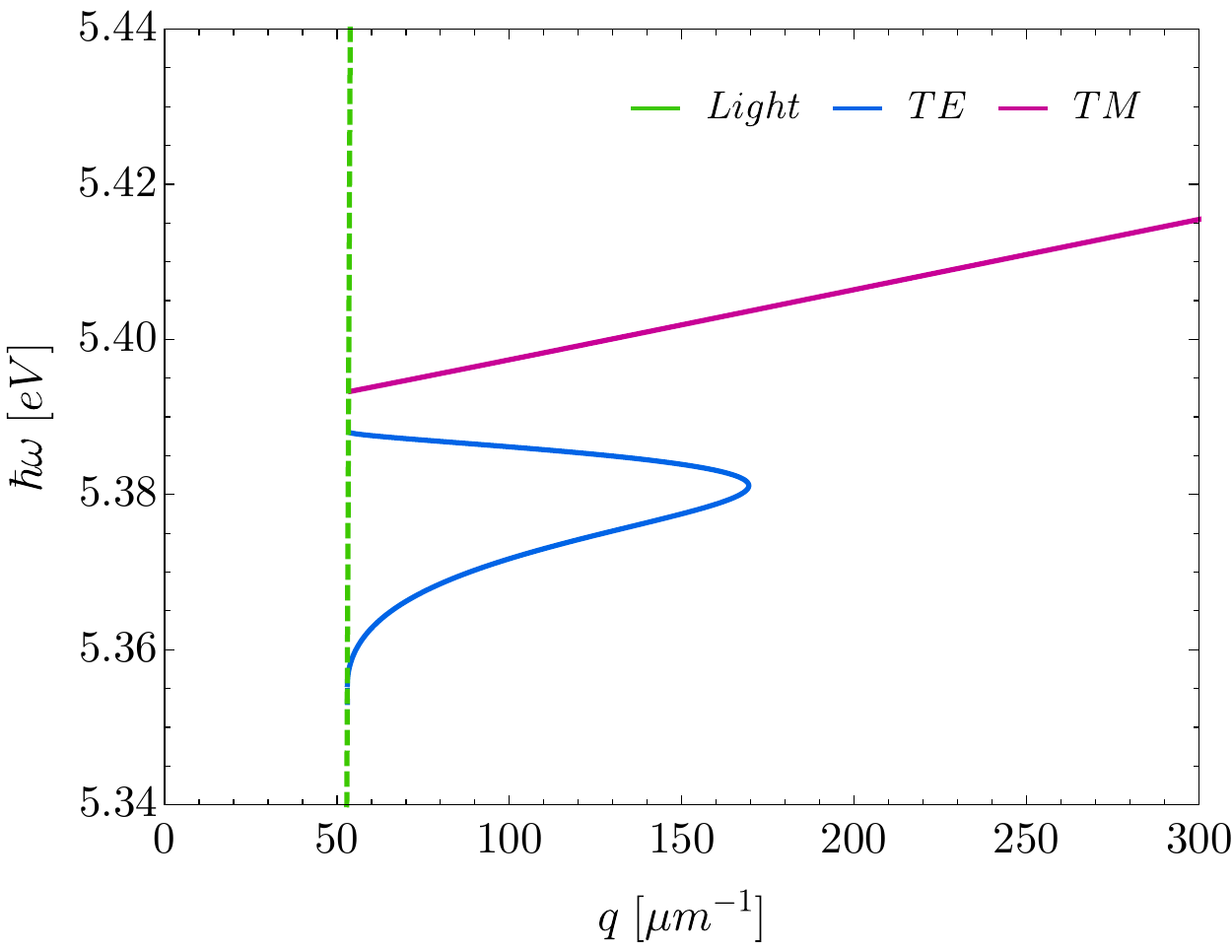}

\caption{Spectrum of the TE and TM exciton-polaritons. We stress that the calculation has taken
into account both the real and the imaginary parts of the optical
conductivity of hBN. The spectrum shown is due to the contribution of the $1s-$exciton alone.
\label{fig:Spectrum-of-the-polaritons}}
\end{figure}

In what concerns their polarization, there can be two different types
of exciton-polaritons: transverse electric (TE) and transverse magnetic (TM). The first kind is possible when
the imaginary part of the excitonic optical conductivity is negative,
whereas the second kind happens in the opposite case, when the imaginary
part of the excitonic optical conductivity is positive. The dispersion relations
of the two kinds of excitons are very different from one other. For TE
exciton-polaritons the implicit expression for the dispersion relation
reads,
\begin{equation}
\kappa_{1}+\kappa_{2}-i\omega\mu_{0}\sigma(\omega)=0\,,\label{eq:TE}
\end{equation}
whereas for TM polarization we have,
\begin{equation}
\frac{\epsilon_{1}}{\kappa_{1}}+\frac{\epsilon_{2}}{\kappa_{2}}+i\frac{\sigma(\omega)}{\epsilon_{0}\omega}=0\,,\label{eq:TM}
\end{equation}
where $\epsilon_{j}$, with $j=1,2$, the relative dielectric permittivity
of the two media cladding the hBN layer; $\kappa_{j}=\sqrt{q^{2}-\epsilon_{j}\omega^{2}/c^{2}}$,
$\omega$ is the frequency of the exciton-polarion electromagnetic
field; $c$ is the speed of light in vacuum; $q$ is the in-plane
wave vector of the polariton; and $\epsilon_{0}$ and $\mu_{0}$ are
the vacuum dielectric permittivity and magnetic permeability, respectively.
The solutions to the two previous equations give the dispersion relations, $\omega(q)$,
for the two possible polarizations. In Fig. \ref{fig:Spectrum-of-the-polaritons},
we depict both the TE and TM dispersion of the polariton.
The lowest branch  refers to the TE polarization
and the other branch to the TM polarization. They both show large
deviations from the light-line, which leads to a strong confinement
of the exciton-polations --- as $q$ grows the exciton-polariton wavelength
$\lambda_{ep}=2\pi/q$ decreases, leading to spatial localization of the
exciton-polariton. A further look at Fig. \ref{fig:Spectrum-of-the-polaritons}
shows that the TM branch has wave numbers that are much larger than those of the TE
branch and, therefore, is prone to a larger degree of confinement.

\begin{figure}
\includegraphics[scale=0.5]{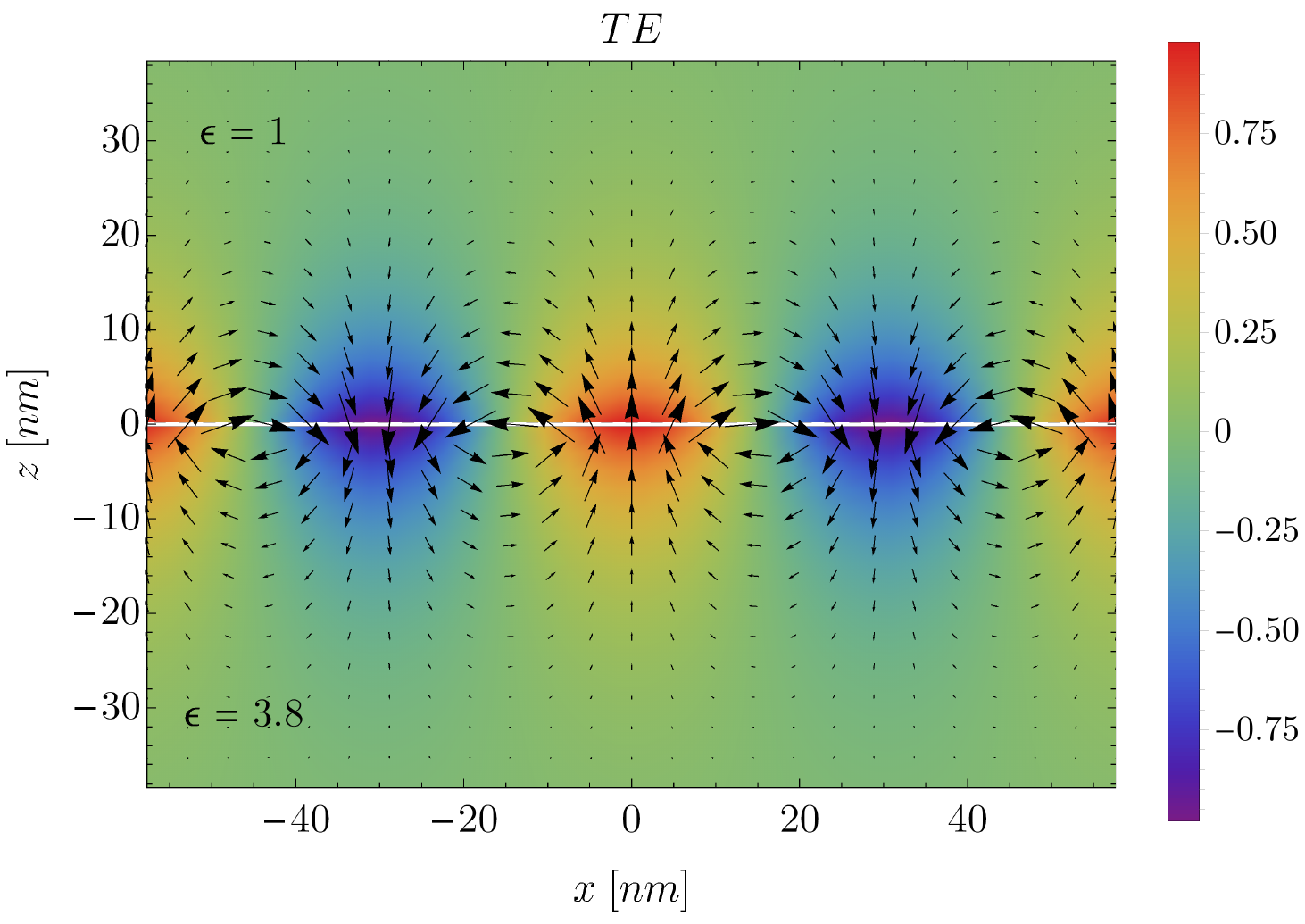}

\includegraphics[scale=0.5]{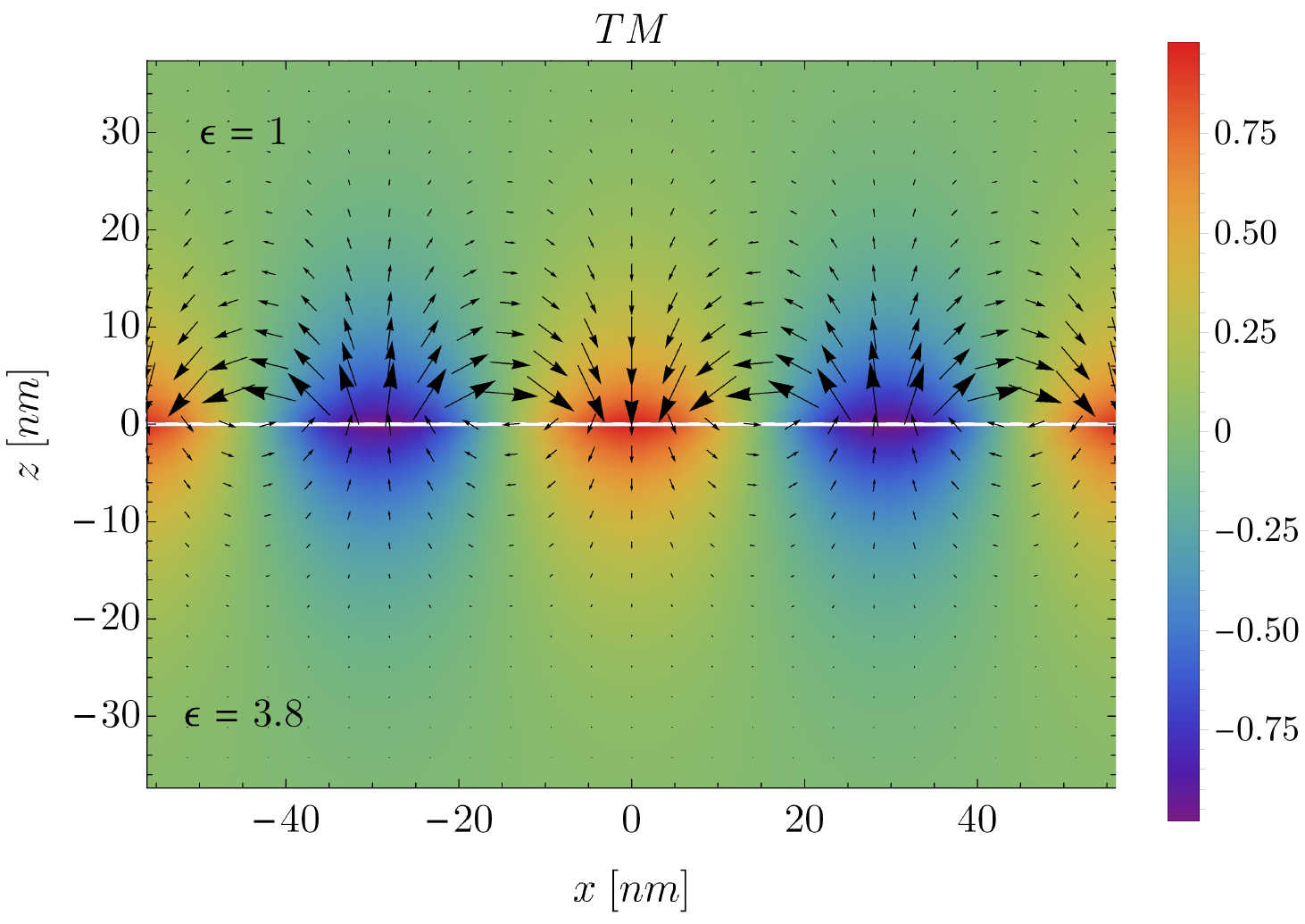}

\caption{Spatial maps of the electromagnetic field of both TE (top) and TM
(bottom) polaritons. For the TE polariton the vector field is that
of the magnetic field and the intensity refers to the single components
of the electric field. For the TM polariton the vector field is that
of the electric field and the intensity refers to the single component
of the magnetic field of the polariton. The parameters used were:
$\epsilon_{1}=1$, $\epsilon_{2}=3.8$,  $q=104\;\mu{\rm m}^{-1}$ and $\hbar\omega=5.386\quad{\rm eV}$
for the TE spatial map (top image), and $q=107\;\mu{\rm m}^{-1}$ and
$\hbar\omega=5.398\quad{\rm eV}$, for the TM spatial map (bottom image).\label{fig:Spatial-maps-of}}

\end{figure}

In Fig. \ref{fig:Spatial-maps-of}, we depict in the top (bottom) panel the electromagnetic
field of the TE (TM) exciton-polaritons. We can see the electromagnetic
field is concentrated within the scale of few nanometers ($\sim10$
nm) for both the TE and TM modes. For the TM mode, one can have a larger
degree of spatial confinement. This difference is explained by the
spectrum shown in Fig. \ref{fig:Spectrum-of-the-polaritons} which
shows that is possible to have TM polarized polaritons that have a larger wave number
than those that are TE polarized. This characteristic --- the strong degree of confinement --- makes the polariton field extremely
sensitive to small changes of the dielectric function of the medium
in contact with the 2D material in the UV range.

\section{The spontaneous radiative decay rate of excitons in hBN\label{sec:The-spontaneous-radiative}}

In this section, we provide the calculation of
the spontaneous radiative decay rate of excitons in hBN. Incidentally, the formula used here
could also be useful for the calculation of this decay recay for the transition-metal dichalcogenides (upon some trivial modifications, that is, considering the spin-orbit band splitting).
For computing the spontaneous radiative decay rate it is convenient
to formulate the problem in second quantization, both for the excitonic
Hamiltonian and for the photon field. For simplicity, we assume that
the hBN layer is in a medium of dielectric permittivity $\epsilon$.
In term of exciton operators $b_{\mathbf{Q},\nu}$ and $b_{\mathbf{Q},\nu}^{\dagger}$
the interaction Hamiltonian of the exciton with the electric field
of the photon reads,

\begin{equation}
H_{int}=e\sqrt{A}\sum_{\nu}\mathbf{E}\cdot\bm{\Phi}_{\nu}^{\ast}b_{\mathbf{0},\nu}^{\dagger}+e\sqrt{A}\sum_{\nu}\mathbf{E}\cdot\bm{\Phi}_{\nu}b_{\mathbf{0},\nu}\,,\label{eq:Hint}
\end{equation}
where $\mathbf{E}$ is the photon field given by,
\begin{equation}
\mathbf{E}(\mathbf{r},t)\approx i\sum_{\mathbf{q}}\sum_{\iota=1}^{2}\sqrt{\frac{\hbar\omega}{2\epsilon\epsilon_{0}V}}[\bm{\epsilon}_{\mathbf{q},\iota}a_{\mathbf{q},\iota}(t)-\bm{\epsilon}_{\mathbf{q},\iota}a_{\mathbf{q},\iota}^{\dagger}(t)]\,,\label{eq:Ephoton}
\end{equation}
the vector $\bm{\Phi}_{\nu}$ reads,
\begin{equation}
\bm{\Phi}_{\nu}=\frac{1}{A}\sum_{\mathbf{k}}\phi_{\nu}(\mathbf{k})\bm{\xi}_{\mathbf{k}vc}\,,
\end{equation}
with $\nu$ representing both the principal and the magnetic quantum
numbers of the exciton, the Berry connection \cite{Blount1962,Ventura2017}
is given by $\bm{\xi}_{\mathbf{k}vc}=-i\langle u_{\mathbf{k},c}\vert\nabla_{\mathbf{k}}\vert u_{\mathbf{k},v}\rangle;$
$V$ is the volume of the box enclosing the photon field; $e$ is
the elementary charge; $\bm{\epsilon}_{\mathbf{q},\iota}$ is a polarization
vector perpendicular to $\mathbf{q}$; $a_{\mathbf{q},\iota}^{\dagger}$
is the creation operator of a photon, while $[a_{\mathbf{q},\iota}^{\dagger}]^{\dagger}=a_{\mathbf{q},\iota}$ is the annihilation operator of a photon.
Note that the approximate equality in Eq. (\ref{eq:Ephoton}) follows from the fact that we are considering the electric field to be homogeneous over the size
of the exciton, that is $\lambda\gg a_{B}$, where $a_{B}$ is the
Bohr radius of the exciton and $\lambda$ the wavelength of the photon.
The Berry connection can be computed using the spinors (\ref{eq:spinor_plus})
and (\ref{eq:spinor_minus}), and leads to:
\begin{equation}
\bm{\xi}_{\mathbf{k}vc}  =-i\frac{v_{F}\hbar e^{-i\theta}}{2M}(\hat{u}_{x}e^{i\theta}-e^{i\theta}i\hat{u}_{y}).
\end{equation}

Having written the interaction Hamiltonian $H_{int}$ it is now possible --- by means of the Fermi golden rule ---
to calculate the spontaneous radiative decay rate of
an exciton in hBN. Since $H_{int}$ depends only on the
exciton operator at $\mathbf{Q}=\mathbf{0}$, momentum conservation
dictates that the in-plane momentum of the emitted photon has to obey
the constraint $(q_{x,}q_{y})=0$. Following this, the radiative decay rate of an exciton with quantum number
$\nu$ and wave vector $\mathbf{Q}$ in hBN reads as

\begin{eqnarray}
\frac{1}{\tau_{r}^{\nu,\mathbf{Q}}}&=&\frac{2\pi}{\hbar}\sum_{\mathbf{q}}\vert\langle GS;1_{\mathbf{q}}\vert H_{int}\vert\nu,\mathbf{Q}\rangle\vert^{2}
\nonumber\\
&\times &
\delta(\hbar\omega-E(\mathbf{Q})-E_{g}-E_{\nu})
\delta_{\mathbf{0},\mathbf{Q}}\delta_{\mathbf{Q},\mathbf{q}_{\parallel}}\,,
\end{eqnarray}
where $E(\mathbf{Q})=\hbar^{2}Q^{2}/(2m_{CM})$ is the energy of the
center of mass of the exciton; $m_{CM}$ is the total mass of the
exciton; and $\vert GS;1_{\mathbf{q}}\rangle$ is the ground state
of the semiconductor with an extra photon of momentum $\mathbf{q}$
in the field. The Kronecker$-\delta$,   $\delta_{\mathbf{Q},\mathbf{q}_{\parallel}}$,
ensures the conservation of the in-plane momentum. We stress that
we are computing the spontaneous emission rate of the exciton since we are assuming that there are no photons in the field prior to the
decay of the exciton. Calculating the matrix element and integrating
over $\mathbf{q}$, finally gives us the radiative decay rate of an exciton in hBN created in a given value,
\begin{equation}
\frac{1}{\tau_{r}^{\nu,\mathbf{0}}}=\alpha\frac{2\pi}{\epsilon\hbar}\frac{v_{F}^{2}\hbar^{2}}{M^{2}}\vert\psi_{\nu}^{m=0}(\mathbf{r}=0)\vert^{2}(E_{g}+E_{\nu})\,,\label{eq:tau}
\end{equation}
where $\psi_{\nu}^{m=0}(\mathbf{r}=0)$ is the exciton wave function
in real space computed at the origin of coordinates with magnetic
quantum number $m=0$; $E_{g}$ is the gap in the spectrum at the
Dirac points, with the correction introduced by the exchange energy;
$\alpha\approx1/137$ is the fine structure constant; and $\nu$ stands
for the principal quantum number. From the parameters of Table \ref{tab:Parameters-used-in} we find that $\hbar/\tau_{r}^{0,\mathbf{0}}\approx 30$
meV for hBN  in vacuum and $\hbar/\tau_{r}^{0,\mathbf{0}}\approx 4.6$ meV for hBN encapsulated in quartz. Although, at the time of writing, the exciton radiative lifetime in hBN has
not yet been measured, the result we obtained is of the order of those found
in the TMD, $\hbar/\tau_{r}\sim4$ meV. The differences should be expected as the TMDs have  smaller gaps than that of the hBN.
Applying Eq. (\ref{eq:tau})
to MoS$_2$ in vacuum, we obtained $\hbar/\tau_{r}\approx4$ meV in agreement with published
results \cite{Maurizia2015} and confirmed experimentally \cite{Kim2018}. This approach  can
also be used to determine the temperature dependence of $1/\tau_{r}^{\nu,\mathbf{0}}$.
For the case where there are photons in the initial state, as it happens
in an experiment illuminated by a continuous-wave laser, the
previous expression is to be modified and acquires a prefactor of the form
$n_{B}(q)+1$, where $n_{B}(q)$ is the Bose-Einstein distribution function of a photon
with wave number $q$. When the temperature rises, the Bose-Einstein distribution function will necessarily
grow and the radiative life time decreases.

\section{Conclusions}

In this paper, we have accurately described the absorption of hBN in
the ultraviolet, in the vicinity of excitonic resonance. We have
developed a semi-analytical method of solving the Wannier equation
based on an expansion of the excitonic wave function in real space
in a basis set of Gaussian functions, a method also used in DFT calculations
of the band structure of solids. The method reduces the calculation
of the energy levels of the exciton and its wave functions to a generalized
eigenvalue problem. The solution of the eigenvalue problem determines
the coefficients of the expansion of the wave function in the Gaussian
basis. Once this problem is solved, the optical properties can be
easily computed for any frequency and non-radiative decay rate. The
same method can be extended to other 2D materials, with the advantage
of being computationally inexpensive. We have also computed the analytical
formula for the spontaneous radiative lifetime of the exciton. It is well known \cite{Kim2018}
that the dielectric susceptibility of a 2D material can be written
as
\begin{equation}
\chi(\omega)=
\frac{c}{d\omega_{0}} \frac{\gamma_{r}}{\omega_{0}-\omega-i\gamma_{nr}/2}\,,\label{eq:chi}
\end{equation}
where $d$ is the thickness of the 2D material, $\omega_{0}$ is the
exciton resonance frequency; $\gamma_{nr}$ the non-radiative decay
rate (encompassing all possible channels), and $\gamma_{r}$ the radiative
decay rate. On the other hand, one can determine the
susceptibility, $\chi(\omega)$,  by means of the Bethe-Salpeter equation. By this calculation with result of Eq. (\ref{eq:chi}), one can obtain $\gamma_{r}$. We have verified (not shown) that the expression for $\gamma_{r}=1/\tau_{r}^{\nu,\mathbf{0}}$ in Eq. (\ref{eq:tau})
agrees with that found from the Bethe-Salpeter equation.

Finally, we would like to stress that the methods developed in this
paper can be applied without restrictions to the study of the optical
properties of other 2D materials. The only necessary conditions are
the existence of a low energy Hamiltonian that describes the single particle
properties, and an energy gap, for implementing the necessary approximations.
Of particular interest is the extension of this method to study anisotropic
2D semiconductors such as phosphorene.

Furthermore --and although we discussed the usefulness of hBN exciton-polaritons
for the detection of cyclic $\beta-$helical peptides-- this same principle
can be applied to other biomolecules that have a strong response in the
ultraviolet, as long as their resonance frequency does not significantly differ from those found on quartz. Another class of these biomolecules are the $\beta-$\emph{Peptoid
Foldamers}, which also present an optical response in the range $[210\,{\rm nm}-240\,{\rm nm}]$
\cite{Laursen2015}. By changing the dielectric substrate, one can tune
the position of the absorption of h-BN across the range of interest
and therefore make from it a viable sensor.

To conclude, we note that our approach can be easily extended to the calculation of nonlinear optical response functions \cite{Hipolito2016} that includes excitonic effects.

\section*{Acknowledgments}
The authors would like to thank André Chaves and Bruno Amorim for
reading the manuscript, making suggestions, and for discussions on the topic
of the paper. 
N.M.R.P. acknowledges support from the European Commission through
the project \textquotedblleft Graphene-Driven Revolutions in ICT and
Beyond\textquotedblright{} (Ref. No. 785219), and the Portuguese Foundation
for Science and Technology (FCT) in the framework of the Strategic
Financing UID/FIS/04650/2019. In addition, N. M. R. P. acknowledges
COMPETE2020, PORTUGAL2020, FEDER and the Portuguese Foundation for
Science and Technology (FCT) through projects PTDC/FIS- NAN/3668/2013
and POCI-01-0145-FEDER-028114, and 
 POCI-01-0145-FEDER-029265 and PTDC/NAN-OPT/29265/2017,
The authors also acknowledge the funding of Fundação
da Ciência e Tecnologia, of COMPETE 2020 program in the FEDER component (European Union),
through the project POCI-01-0145-FEDER-02888.

\appendix

\section{Numerical solution of the 2D Coulomb problem\label{sec:App Numerical Sol}}

In this Appendix, we benchmark the numerical method against the known
analytical solution of the attractive 2D Coulomb problem. We also
give the Hamiltonian and superposition kernels.

\subsection{The Hamiltonian and superposition kernels, and numerical results
for the Coulomb potential}

The numerical part of the method requires the calculation of the Hamiltonian
and superposition kernels. In the case of the Coulomb potential, these
have a simple analytical form for a given quantum number $m$. The
superposition kernel reads
\begin{equation}
S(m,\zeta_{i},\zeta_{j})=\pi(\zeta_{i}+\zeta_{j})^{-1-|m|}\Gamma(|m|+1)\,,
\end{equation}
while the kinetic kernel is given by
\begin{equation}
K^{kin}(m,\zeta_{i},\zeta_{j})=\frac{2\pi\hbar^{2}}{\mu}\zeta_{i}\zeta_{j}(\zeta_{i}+\zeta_{j})^{-\left|m\right|-2}\Gamma(\left|m\right|+2)\,,
\end{equation}
where $\mu$ is the reduced mass of the exciton. The Coulomb potential
kernel reads
\begin{equation}
K^{C}(m,\zeta_{i},\zeta_{j})=-\frac{e^{2}}{4\epsilon_{0}}(\zeta_{i}+\zeta_{j})^{-|m|-1/2}\Gamma\left(|m|+\frac{1}{2}\right)\,.
\end{equation}
With these three analytical expressions we can construct the $H$ and
$S$ matrices, and subsequently solve the generalized eigenproblem ---
obtain the energy spectrum and the wave functions.

\begin{figure}
\includegraphics[scale=0.5]{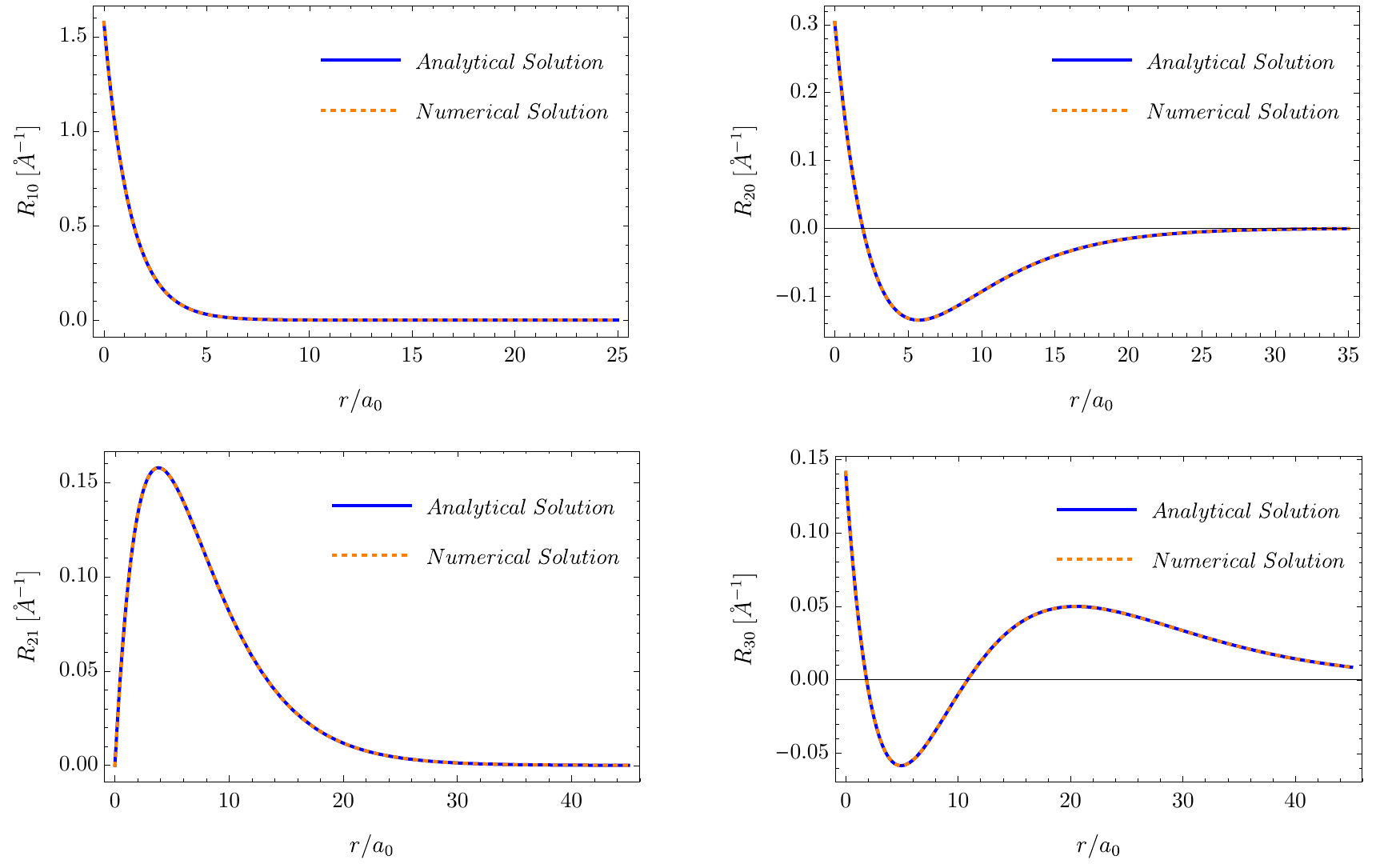}

\caption{\label{fig:AnalyticalVsNumericalWF}Comparison of numerical and analytical
solutions for the Hydrogen radial wavefunctions $R_{10}$, $R_{20}$,
$R_{21}$, and $R_{30}$, with $a_{0}$  the Bohr radius. We can
see an excellent agreement between the two sets of wave functions,
to the point that the overlaid plots are almost indistinguishable.}
\end{figure}

The analytical energy spectrum and wave functions of the attractive
2D Coulomb potential are well known \cite{Yang1991}, and are given by

\begin{equation}
E_{n}=-\frac{m_{0}e^{4}Z^{2}}{2(n-1/2)^{2}(4\pi\epsilon_{0})^{2}\hbar^{2}},\quad\quad n=1,2,3,...\,,\label{eq:En2D}
\end{equation}
for the energy spectrum and 
\begin{eqnarray}
R_{nm}(r)&=&A_{n}(\beta_{n}r)^{|m|}e^{-\beta_{n}r/2}\times
\nonumber\\
&&F_{1}^{1}(-n+|m|+1,2|m|+1,\beta_{n}r)\,,
\end{eqnarray}
for the radial wave function, with 
\begin{equation}
A_{n}=\frac{\beta_{n}}{(2|m|)!}\bigg[\frac{(n+|m|-1)!}{(2n-1)(n-|m|-1)!}\bigg]^{1/2}\,,
\end{equation}
 and 
\begin{equation}
\beta_{n}=\Bigg(\frac{2m_{0}e^{2}}{\hbar^{2}4\pi\epsilon_{0}}\Bigg)\frac{1}{n-1/2}\,.
\end{equation}
In Table \ref{tab:AnalyticalVsNumerical Energy Spectrum} we compare
the numerical determined energy spectrum with those given by Eq. (\ref{eq:En2D}).
We see that the agreement is complete. In Fig. \ref{fig:AnalyticalVsNumericalWF}, the
the numerical and the analytical wave functions are compared and one can see that the two solutions are almost entirely superposed. As for the energy spectrum, again the agreement between the numerical and the analytical results is nearly perfect.

\begin{table*}
\centering{}%
\begin{tabular}{|c|c|c|c|c|}
\hline
principal q.n. & $n=1$ & $n=2$ & $n=3$ & $n=4$\tabularnewline
\hline
\emph{Numerical} & -2.836 eV & -0.315 eV & -0.113 eV & -0.0579 eV
\\
\hline
\emph{Analytical} & -2.834 eV & -0.315 eV & -0.113 eV & -0.0578 eV
\\
\hline
\end{tabular}
\caption{\label{tab:AnalyticalVsNumerical Energy Spectrum}Comparison of numerical
and analytical values of 2D hydrogen energy levels, for $m=0$, and dielectric constant $\epsilon=2.4$. Although the table
only shows the energies obtained up to the principal quantum number (q.n.) $n=4$, the numerical results
are in agreement with the analytical ones up to $n=6$ with differences
smaller than $1\%$. The calculation used $100$ Gaussian functions. The effective mass is $m^\ast=0.6m_0$.}
\end{table*}

\subsection{The interacting kernel for the Rytova-Keldysh potential}

As in the Coulomb problem, it is also possible to obtain an analytical
expression for the Hamiltonian and superposition kernels of the attractive
Rytova-Keldysh potential. Obviously, the kinetic energy kernel and
the superposition kernel are the same as those given in the previous
section. The Rytova-Keldysh energy kernel is given by 
\begin{eqnarray}
K^{RK}(m,\zeta_{i},\zeta_{j})  =\frac{e^{2}}{8r_{0}^{2}\epsilon_{0}}(\zeta_{i}+\zeta_{j})^{-|m|-3/2}
\times
\nonumber\\
\left[-2\epsilon_{1}\Gamma\left(\frac{3}{2}+|m|\right){}_{2}F_{2}\left(1,\frac{3}{2}+|m|;\frac{3}{2},\frac{3}{2};-\frac{\epsilon_{1}}{4r_{0}^{2}(\zeta_{i}+\zeta_{j})}\right)\right.
\nonumber\\
\left.
  +\pi r_{0}\sqrt{\zeta_{i}+\zeta_{j}}G_{11}^{21}\bigg(\frac{\epsilon_{1}^{2}}{4r_{0}^{2}(\zeta_{i}+\zeta_{j})}\bigg|
\begin{array}{c}

{ {-|m|;-1/2}}\\
 { {0,0;-1/2}}
\end{array} 
\bigg)\right]\,,
\end{eqnarray}
where $_{p}F_{q}$ is the generalized hypergeometric function, $G_{pq}^{mn}$
is the Meijer-G function, and $\epsilon_{1}$ is the average dielectric
function of the substrate and capping layer.

\end{document}